\documentclass[reprint,aps,pra,amsmath,amssymb,superscriptaddress,floatfix,%
  notitlepage]{revtex4-1}
\usepackage[pdftex]{graphicx} 
\newcommand{\Sys}{\mathcal{S}}

\newcommand{\ket}[1]{\vert #1 \rangle}

\newcommand{\op}[2]{\vert#1\rangle\langle#2\vert}

\newcommand{\abs}[1]{\vert #1 \vert}

\newcommand{\Heff}{H_{\text{eff}}}
\newcommand{\norm}[1]{\vert\vert #1\vert\vert}

\newcommand{\mean}[1]{\mathsf{E}\left[ #1 \right]}

\begin{document}
\title{Non-Markovian waiting time distribution for quantum jumps in open systems}

\author{Kimmo Luoma}
\email{ktluom@utu.fi}
\affiliation{Turku Centre for Quantum Physics, Department of Physics and 
Astronomy, University of Turku, FI-20014, Turun yliopisto, Finland}

\author{Kari H\"ark\"onen}
\affiliation{Max-Planck-Institute f\"ur Physik komplexer Systeme,
N\"othnitzner stra\ss{}e 38, D-01187 Dresden, Germany}

\author{Sabrina Maniscalco}
\affiliation{Turku Centre for Quantum Physics, Department of Physics and 
Astronomy, University of Turku, FI-20014, Turun yliopisto, Finland}
\affiliation{SUPA, EPS/Physics, Heriot-Watt University, Edinburgh, EH14 4AS, 
  United Kingdom}

\author{Kalle-Antti Suominen}
\affiliation{Turku Centre for Quantum Physics, Department of Physics and 
Astronomy, University of Turku, FI-20014, Turun yliopisto, Finland}


\author{Jyrki Piilo}
\affiliation{Turku Centre for Quantum Physics, Department of Physics and 
Astronomy, University of Turku, FI-20014, Turun yliopisto, Finland}

\date{\today}

\begin{abstract}
Simulation methods based on stochastic realizations of state vector evolutions 
are commonly used tools to solve open quantum system dynamics, both in the 
Markovian  and non-Markovian regime.  Here, we address the question of 
waiting time distribution (WTD) of 
quantum jumps  for non-Markovian systems. We generalize Markovian quantum 
trajectory methods in the sense of deriving an exact analytical WTD for 
non-Markovian quantum dynamics and show explicitly how to construct this 
distribution for certain commonly used quantum optical systems. 
\end{abstract}

\pacs{03.65.Yz, 42.50.Lc}
\maketitle
\section{Introduction}
An open quantum system interacting with its environment undergoes non-unitary 
evolution and typically loses its quantum properties, such as entanglement, 
due to decoherence~\cite{Breuer2007}. Whilst the theory of Markovian dynamics 
in terms of semigroups and completely positive trace preserving maps is fairly 
well understood since the pioneering work of Lindblad, Gorini, Kossakowski, 
and Sudarshan~\cite{Lindblad1976,gorini:821},  non-Markovian quantum dynamics 
displaying memory effects has become under active study during the recent 
years.  The advances here include the development of simulation 
schemes~\cite{PhysRevA.50.3650,Breuer1999,PhysRevLett.82.1801,PhysRevA.64.053813,PhysRevA.66.012108,PhysRevE.66.037701,PhysRevA.69.042107,PhysRevA.69.052104,Breuer2004,Piilo2008,0295-5075-85-5-50004,PhysRevLett.105.240403,PhysRevA.84.032113}, the limits for the existence of 
physically valid dynamical 
maps~\cite{PhysRevLett.102.100402}, the discussion about the applicability 
of different types 
of master equations~\cite{PhysRevLett.104.070406}, the very definition and 
quantification of 
quantum non-Markovianity~\cite{PhysRevLett.101.150402,Breuer2009a,PhysRevLett.105.050403}, 
and the role of initial correlations between the system and its environment~\cite{PhysRevA.70.052110,Shaji200548,PhysRevA.77.042113,1751-8121-41-20-205301,PhysRevA.82.012341,0295-5075-92-6-60010}. 
Moreover, it is also possible to control and quantify experimentally the 
non-Markovian features of quantum dynamics~\cite{NonMarkovMeasureNature2011,0295-5075-97-1-10002} and the influence of 
initial system-environment correlations~\cite{PhysRevA.83.064102,PhysRevA.84.032112}. Subsequently, this progress 
allows to look for ways how non-Markovian features with memory effects can be 
exploited for quantum information processing~\cite{PhysRevA.83.042321}, and for 
quantum control and engineering tasks~\cite{2011arXiv1106.2841H,Chin2011}.

Here, our focus is on fundamental aspects of non-Markovianity and, in 
particular, on the jumplike stochastic unravellings, or simulation schemes, for 
open system dynamics~\cite{0295-5075-85-5-50004,Piilo2009,Piilo2008,Breuer2004,Breuer1999,PhysRevA.69.052104,PhysRevE.66.037701,PhysRevA.69.042107,Harkonen2010}. For Markovian systems, 
some of the most popular stochastic schemes include the Monte Carlo Wave
Function (MCWF) \cite{Dalibard1992,Molmer1993} and 
Quantum Trajectory (QT) \cite{Dum1992,PhysRevA.46.4382,Carmichael2008} methods. In both of 
these methods the time evolution of a singe realization consists of periods of 
continuous 
deterministic evolution interrupted by stochastic jumps, i.e.,
both methods simulate a piecewise deterministic stochastic process (PDP). In 
MCWF method, the time evolution of a single realization progresses in a 
stepwise  
fashion, e.g. during each time step we decide whether the realization 
evolves deterministically or jumps. The mean time evolution of the ensemble of 
realizations, 
over small time increments, matches with the solution of a Markovian master 
equation for the density matrix (for the first order in time increment).
The central concept for the QT methods, in turn, is the waiting time 
distribution (WTD).
The random jump time of the realization can be sampled from the 
WTD and the state vector is directly evolved 
deterministically till this point.
Solution to the Markovian master equation is formed 
from the weighted average over all possible stochastic evolutions 
that realizations might take. Generally speaking, the MCWF method exploits 
the increments of the WTD while the QT uses the full exact form of the WTD. 

A few years ago MCWF was generalized to non-Markovian region by
Non-Markovian Quantum Jump method (NMQJ)~\cite{0295-5075-85-5-50004,Piilo2009,Piilo2008}. In NMQJ, evolution
of the ensemble average over a time step $\delta t$, matches 
with the solution given by local in time master equation with possibly
temporarily negative rates. The central ingredient of the NMQJ method is a 
quantum jump which can restore coherence, e.g., by returning the stochastic 
realization to the superposition which was destroyed earlier. Formally, the 
probability of the reverse jump can be calculated using the concept of 
positive definite jump probability density \cite{Harkonen2010}. However, to 
the best of our knowledge, the QT methods -- without using the auxiliary 
extensions of the state space of an open quantum system -- have not yet been 
extended to 
the non-Markovian region. The main obstacle here has been the fact that the 
WTD for non-Markovian systems, when calculated along the Markovian line of 
reasoning, displays oscillations which render its physical meaning invalid 
and prevent the technical implementation of the simulations, whilst the 
mathematical calculation of the WTD still is, in some sense, correct.    

With the help of the insight provided by the NMQJ method and the concept of 
positive definite jump probability density, we derive a general analytical 
form of the waiting distribution, which is both physically and mathematically 
correct for non-Markovian quantum dynamics. This is the main result of our
paper. 
We thereby generalize the QT formalism into the non-Markovian regime and 
show explicitly how to construct the WTD for some commonly used quantum 
optical systems. It is worth keeping in mind here that, as already featured in 
the NMQJ method, the stochastic realizations depend on each other as a 
consequence of the memory effects. Moreover, it has been recently argued 
that non-Markovian unravelings can not be interpreted as stemming 
from continuous measurement of the environment~\cite{PhysRevLett.101.140401} 
despite of some attempts in that direction~\cite{PhysRevA.85.034101,PhysRevLett.100.080401}.
It seems to us, that the functional form of the derived non-Markovian WTD 
indicates the former choice of answers.

The structure of the paper is the following. In Sec.~\ref{sec:PIECEWISE} we 
introduce the PDP corresponding to the NMQJ method and most importantly the
positive definite jump probability density.
In Sec.~\ref{sec:WTD}, we give the general form of the waiting 
time distribution  and connect it 
to the PDP defined in Sec.~\ref{sec:PIECEWISE}.
In Sec.~\ref{sec:EXAMPLES} we present some quantum optical 
examples that illustrate 
the general construction of the WTD and the effects of non-Markovianity,  
in Sec.~\ref{sec:DISCUSS} we present some further discussion about our results 
and conclude in Sec.~\ref{sec:CONCLUSIONS}. 

\section{\label{sec:PIECEWISE} Piecewise deterministic process for non-Markovian system}
In this section we formulate non-Markovian piecewise
deterministic process for pure states $\psi$ \cite{Piilo2008,0295-5075-85-5-50004,Harkonen2010}.
Reduced state of the system, $\rho$, is obtained as an ensemble average
\begin{align}
  \label{eq:5}
  \rho(t)=&\mean{\op{\psi}{\psi}}=
  \int\text{d}\psi P[\psi,t]\op{\psi}{\psi},
\end{align}
where $\text{d}\psi=D\psi D\psi^*$ is a
singular volume element of the Hilbert space of the system and 
$P[\psi,t]$ is time dependent, phase invariant probability density 
functional concentrated 
on the surface of a unit sphere ($\norm{\psi}=1$).
$\rho(t)$ solves also the following
time convolutionless (TCL) master equation \cite{Breuer2007}
\begin{align}\label{eq:Master}
\dot{\rho}(t)&=-i\hbar^{-1}\left[H_\Sys(t),\rho(t)\right]\notag\\
&+
\sum_i\Delta_{i}(t)\left(C_i\rho(t)C_i^\dagger-
\frac{1}{2}\left\{\rho(t),C_i^\dagger C_i\right\}\right)\notag\\
&=-i\hbar^{-1}\left[H_\Sys(t),\rho(t)\right]\notag\\
&+
\sum_j\Delta_{j}^+(t)\left(C_j\rho(t)C_j^\dagger-
\frac{1}{2}\left\{\rho(t),C_j^\dagger C_j\right\}\right)\notag\\
&-
\sum_k\Delta_{k}^-(t)\left(C_k\rho(t)C_k^\dagger-
\frac{1}{2}\left\{\rho(t),C_k^\dagger C_k\right\}\right),
\end{align}
where $\Delta_i(t)$ is time dependent decay rate. After the second equality
sign, we have split the decay rates into two components 
$\Delta_j^\pm(t)=\left(\abs{\Delta_j(t)}\pm\Delta_j(t)\right)/2$ to 
account better for the overall sign of the decay rate \cite{0295-5075-85-5-50004}.
However, note that $\Delta_j^\pm(t)$ are non-negative for all times $t$. 
From now on we assume that $\hbar=1$. 
Operators $C_j$ are called 
jump operators and we make a simplifying assumption that 
they are time invariant. 
Unnormalized 
states  are labeled with $\tilde\psi$ and normalized with $\psi$. 
We formulate the process for pure initial states only, since mixedness adds no 
novelty here. 

Between two subsequent jumps at times $T$ and $t=T+\tau$ ($\tau>0$),
pure states evolve deterministically according to an  
effective non-Hermitian Hamiltonian
\begin{align}\label{eq:HEFF}
\Heff(t)=& H_\Sys(t)-\frac{i}{2}\sum_j\Delta_j(t)C_j^\dagger C_j,
\end{align}
such that state $\psi(t)$ is expressed as
\begin{align}\label{eq:8}
  \psi(t)=\psi(T+\tau)=\frac{\tilde\psi_T(\tau)}
         {\norm{\tilde\psi_T(\tau)}},
\end{align} 
where $\tilde\psi_T(\tau)$ satisfies the Schr\"odinger equation
$\dot{\tilde\psi}_T(\tau)=-i\Heff(T+\tau)\tilde\psi_T(\tau)$ with 
the initial condition $\tilde\psi_T(0)=\psi(T)$. 

Discontinuous part of the process consists of
jumps between different pure states. Given that the process is
in pure state $\psi$,
conditional jump probability density from a source state $\psi$ to a target 
state $\phi$ using a channel $k$ during a time 
interval $\left[T,T+\delta t\right]$ is
\cite{Harkonen2010}
\begin{align}\label{eq:conditional_jump_density}
  p_k[\phi|\psi,T] =& \delta t \Delta_k^+(T)\norm{C_k\psi(T)}^2
  \delta\left[\phi(T)-\frac{C_k\psi(T)}{\norm{C_k\psi(T)}}\right]\notag\\
  &+
  \delta t \Delta_k^-(T)\frac{P\left[\phi,T\right]}{P\left[\psi,T\right]}
  \norm{C_k\phi(T)}^2\notag\\
  &\times\delta\left(\psi(T)-
  \frac{C_k\phi(T)}{\norm{C_k\phi(T)}}\right).
\end{align}
Above, $\delta$-functional satisfies 
$\int\text{d}\phi\,\delta(\psi-\phi)F[\phi]=F[\psi]$ where 
$F$ is an arbitrary smooth functional. The 
$\delta$-functionals in Eq.~\eqref{eq:conditional_jump_density} give
temporal channel-wise stochastic connection between different regions 
of projective Hilbert space (global phase of the states 
is irrelevant). Connection of the positive part
(i.e. part proportional to $\Delta_k^+$)
is of one-to-one type: $\psi\to\frac{C_k\psi}{\norm{C_k\psi}}$, 
which corresponds to Markovian quantum jumps.
Interestingly,
connection of the negative part is one-to-many type:
Each source state $\psi$ may jump to one of the states 
$\left\{\phi\right\}$ that satisfy
$\psi=\frac{C_k\phi}{\norm{C_k\phi}}$ provided that the corresponding jump probability is nonzero. It follows that the connection
provided by the negative part requires the knowledge of the 
different states 
in the pure state 
decomposition of $\rho$, 
since the range of the one-to-many mapping is not obtainable from
the structure of Eq.~\eqref{eq:Master}.
To summarize, a negative channel induces a one-to-many mapping 
for the pure states and therefore, in general, 
one decay channel connects several different 
regions of the projective Hilbert space stochastically.

Next we sketch the stepwise progression 
of the PDP, more details may be found in Refs.~\cite{Piilo2009,Harkonen2010}.
During an interval $I=[T+\tau,T+\tau+\delta t]$, 
a realization of the process in state $\psi$  may either jump or evolve 
deterministically. 
The total jump rate away from state $\psi$ during the interval $I$ is
the total jump probability to any other state 
via  any channel divided by the length of the interval
\begin{align}\label{eq:tot_rate}
  \Gamma\left[\psi,T+\tau\right]=&\frac{1}{\delta t}
  \int\text{d}\phi\,\sum_k p_k[\phi|\psi,T+\tau].
\end{align}
Therefore, with probability $1-\Gamma\left[\psi,T+\tau\right]\delta t$,
the realization does not jump away from state $\psi$ but 
evolves deterministically. Deterministic evolution is governed by 
the Schr\"odinger equation and Eq.~\eqref{eq:HEFF}.
With probability $\Gamma\left[\psi,T+\tau\right]\delta t$ the realization jumps;
the target state of the jump is chosen from the probability distribution 
$\frac{p_k[\phi|\psi,T+\tau]}{\Gamma[\psi,T+\tau]\delta t}$.
After the stochastic evolution of the ensemble over a small
time step, the average over the ensemble  
provides us 
Eq.~\eqref{eq:Master} for the first order in $\delta t$.

\section{\label{sec:WTD} Waiting time distribution for non-Markovian system}
In this section we derive the general form of the waiting time distribution,
which is valid also for non-Markovian systems, starting from
the positive definite jump probability density. We also provide a 
formula for estimating the WTD from a sample of realizations.

\subsection{Analytical WTD}
\label{sec:analytical-wtd}
By definition, the waiting time distribution $F(\tau|\psi,T)$ 
is a conditional probability distribution function which
gives the probability for the next jump to occur during a 
time interval $\left[T,T+\tau\right]$ conditioned on that 
at time $T$ the state of the realization is known to be  
$\psi$ \cite{Breuer2007}. 

Probability for a jump to occur during a short time interval  
$I=\left[T+\tau,T+\tau+\delta t\right]$ away from state $\psi$ 
is then $\delta F(\tau|\psi,T)\equiv F(\tau+\delta t|\psi,T)-F(\tau|\psi,T)$,
which is equal to the probability of having no jumps before 
$T+\tau$ and a jump during the 
following $\delta t$, i.e. $\delta F(\tau|\psi,T)=
(1-F(\tau|\psi,T))\Gamma[\psi,T+\tau]\delta t$. 
Dividing both sides by $\delta t$  and taking the limit 
$\delta t\to 0$ we obtain the 
following differential equation that every valid
WTD must satisfy~\cite{Breuer2007}
\begin{align}\label{eq:WTD_ODE}
  \frac{\text{d}}{\text{d}\tau}F(\tau|\psi,T)=& 
  \left[1-F(\tau|\psi,T)\right]\Gamma\left[\psi,T+\tau\right].
\end{align}
This can be solved formally with an initial condition  
$F(0|\psi,T)=0$, such that  
\begin{align}\label{eq:WTD_IMPLICIT}
  F(\tau|\psi,T)=&1-\exp\left\{-\int_T^{T+\tau}\text{d}s\Gamma[\psi,s]\right\}.
\end{align}
Then, by using Eqs.~\eqref{eq:conditional_jump_density},~\eqref{eq:tot_rate},
and Eq.~\eqref{eq:WTD_IMPLICIT}
we can obtain the following form for 
the generic WTD corresponding to Eq.~\eqref{eq:Master}:
\begin{align}\label{eq:WTD_GEN}
  F(\tau|\psi,T)=&1-\exp\bigg\{-\int_T^{T+\tau}\text{d}s\,
  \int\text{d}\phi\notag\\
  &\times\sum_k\bigg(\Delta_k^+(s)\norm{C_k\psi(s)}^2
  \delta\left[\phi(s)-\frac{C_k\psi(s)}{\norm{C_k\psi(s)}}\right]\notag
  \\
  &+\Delta_k^-(s)\frac{P\left[\phi,s\right]}{P\left[\psi,s\right]}
  \norm{C_k\phi(s)}^2\notag\\
  &\times\delta\left[\psi(s)-\frac{C_k\phi(s)}{\norm{C_k\phi(s)}}\right]
  \bigg)\bigg\}.
\end{align}
Terms proportional to $\Delta_k^+$ depend 
only on the state of the particular realization, 
its deterministic time evolution
and the quantities obtainable from Eq.~\eqref{eq:Master}.
Terms proportional to $\Delta_k^-$ are more complicated, 
since they depend on the probability functionals and on the deterministic 
time evolution of other states to which the realization might jump 
via a channel-wise one-to-many 
mapping.

Random waiting time $\tau^\star$ is sampled from waiting time distribution
by comparing a random number $\eta$ to the WTD: 
$\tau^\star(\eta)=\min\{\tau|F(\tau|\psi,T)>\eta\}$ 
\cite{Breuer2007}. 
Probabilities $P[\psi,s]$ appear on the right hand side of
Eq.~\eqref{eq:WTD_GEN} and they are modified each time a jump occurs in the 
ensemble.

When all decay rates $\Delta_i(t)$ for all times $t$ are non-negative  
in Eq.~\eqref{eq:Master},    
the total jump rate away from pure state $\psi$ is 
$\Gamma[\psi,t]=\sum_k\Delta_k(t)\norm{C_k\psi(t)}^2$. 
Inserting this into Eq.~\eqref{eq:WTD_GEN} and taking into
account the deterministic evolution of $\psi$, we obtain the following 
familiar Markovian limit for the WTD
\cite{Breuer2007,Dum1992,PhysRevA.46.4382,PhysRevA.46.4363}:
\begin{align}\label{eq:WTD_MARKOV}
  F(\tau|\psi,T)=&\frac{\norm{\tilde\psi_T(0)}^2-\norm{\tilde\psi_T(\tau)}^2}
  {\norm{\tilde\psi_T(0)}^2}.
\end{align}
Details of the derivation of the Markovian limit can be found in the 
Appendix~\ref{sec:markovian-limit}.

\subsection{Estimation of WTD}
\label{sec:estimation-wtd}
We assume that the reduced state of a 
non-Markovian open quantum system can be 
expressed at all times, as a linear combination
of a finite number of, in general, non-orthogonal pure state projectors.
Then we can write Eq.~\eqref{eq:5} as 
\begin{align}\label{eq:decomp}
  \rho(t)=\sum_\alpha P_\alpha(t)\op{\psi^\alpha}{\psi^\alpha}.
\end{align}
Assume that  we have a sample of $N_S$ realizations  
from PDP in Sec.~\ref{sec:PIECEWISE} over a time interval $[t_0,t_s]$ divided
into $N_t$ time steps.
The samples are collected to an $N_t\times N_S$ matrix  
$\mathbf{M}$ where the element $\mathbf{M}_{i,j}=\beta$ means that 
a realization $j$ is in state $\psi^\beta$ at time $t_i=(i-1)\delta t+t_0$.
Set of column indices $I_i^\beta$ of row $i$ of $\mathbf{M}$, give the 
indices of the realizations which are in 
state $\beta$ at time $t_i$. Hence each set $I_i^\beta$ has 
$N_S$ elements, where the $k$th element is $1$ if realization $k$ is in state 
$\beta$ at time $t_i$, otherwise the $k$th element is 0. 
$|I_i^\beta|=\sum_{k=1}^{N_S}(I_i^\beta)_k$ is the total number of realizations 
in state $\beta$ at time $i$.

If we know that the realization $r$ is in state $\alpha$ at 
time $t_i$, then the discrete sample estimate for the 
probability to jump away from state $\alpha$ during the discrete
time interval  $[t_i,t_j]$ is
\begin{align}\label{eq:WTD_EST}
  W_r(t_k|t_i,\alpha)=&1-\sum_{l=i+1}^k\frac{|I_{l-1}^\alpha\cap I_{l}^\alpha|}
  {|I_i^\alpha|}.
\end{align}
Naturally we have that $W_r(t_i|t_i,\alpha)=0$. 
The meaning of this equation is 
that the intersection of two sets consists of the indices 
of those realizations that 
were in a state $\alpha$ at the previous time and are still there 
at the present time. The number of such realizations is divided by the 
number of the realizations in $\alpha$ at time $t_i$ (beginning of the 
time interval). 

\section{Construction of WTD for quantum optical systems}
\label{sec:EXAMPLES}
\begin{figure}
  \includegraphics[width=8cm]{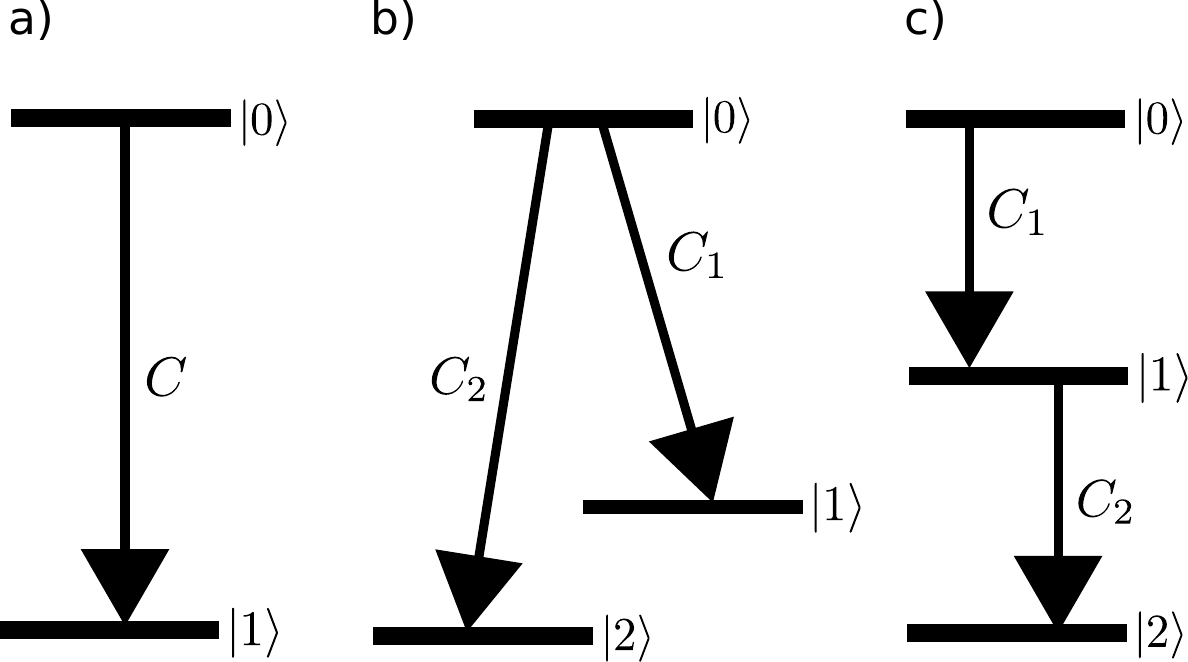}
  \caption{\label{fig:systems}(Color online)
  Schematic figures of the different systems. 
  a) Two level system, b) $\Lambda$ system and c) Ladder system. }
\end{figure}
In this section we construct the waiting time distribution
explicitly for a few simple quantum optical systems interacting with a leaky
cavity mode. In Fig.~\ref{fig:systems} we have presented schematically 
the different systems that we shall study.
\subsection{Two level system}\label{sec:TLA}
Orthonormal basis for the Hilbert space of the system  is 
$\left\{\ket{0},\ket{1}\right\}$, where 
$\ket{1}$ is the ground state and $\ket{0}$ is the excited state of a 
two level atom (TLA).
(see Fig.~\ref{fig:systems}a)).
The initial state is $\psi^0(t_0)=c_0(t_0)\ket{0}+c_1(t_0)\ket{1}$, which
is the only state with non-trivial deterministic evolution.
The state of the system is decomposed for all times $t$
as $\rho(t)=P_0(t)\op{\psi^0}{\psi^0}+P_1(t)\op{\psi^1}{\psi^1}$,
where $\psi^1$ is the ground state.
The detailed description of the system is given in Appendix~\ref{app:TLA}.

\begin{figure}
  \includegraphics{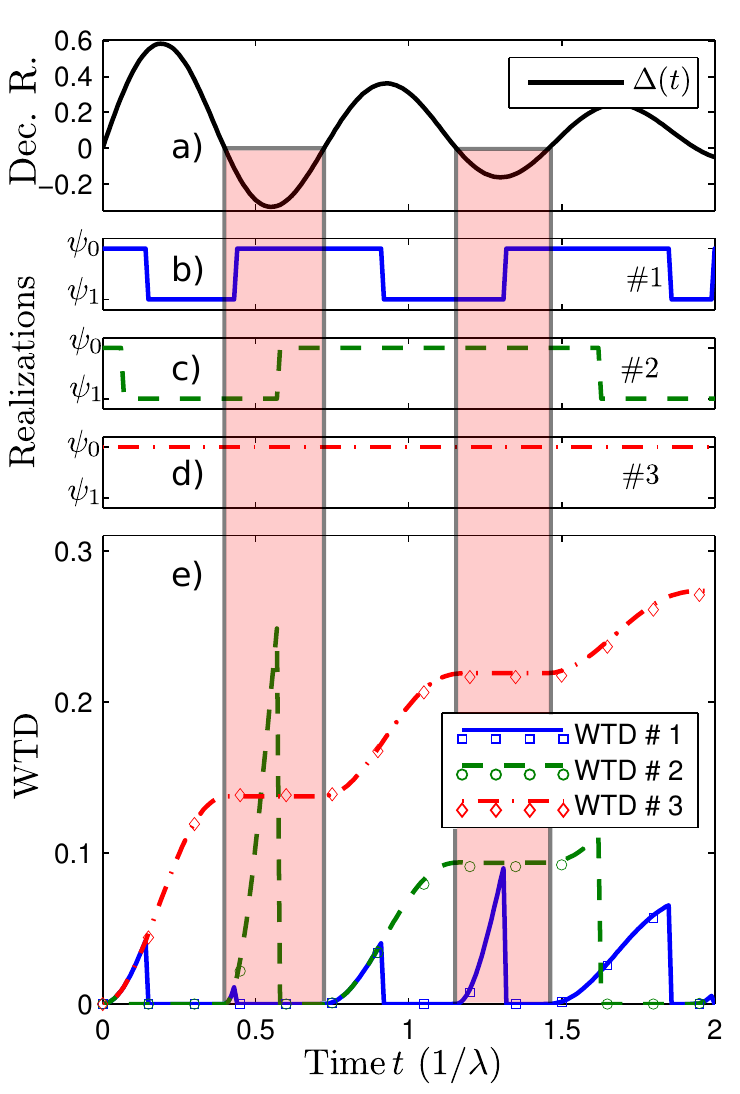}
  \caption{\label{fig:TLA}(Color online)
  Initial state is  $\ket{\psi^0(0)}=\ket{0}$ and parameters are $\gamma_0=5\lambda$, 
  $\delta=8\lambda$ and sample size was $10^5$. In panel $a)$
  we have the decay rate, in $b)-d)$ we have three 
  different realizations. In panel $e)$ we have 
  WTDs for the realizations. Line style and color coding match with
  the sample realization. Lines are for exact numerical solution and markers
  for sample estimate.}
\end{figure}

The non-Hermitian Hamiltonian generating the deterministic pieces of the 
time evolution is obtained from Eq.~\eqref{eq:HEFF} (see  
details in the Appendix~\ref{app:TLA}).
The total rate away from the deterministic state $\psi^0(t)$ is 
\begin{align}\label{eq:TLA_RATEPOS}
  \Gamma[\psi^0,t]=
  \begin{cases}
    \Delta(t)\norm{C\psi^0(t)}^2,&\,\Delta(t)\geq 0,\\
    0, &\,\Delta(t)<0,
  \end{cases}
\end{align}
and the total rate away from the state $\psi^1$ is
\begin{align}\label{eq:TLA_RATENEG}
  \Gamma[\psi^1,t]=
  \begin{cases}
    0, &\,\Delta(t) \geq 0,\\
    \abs{\Delta(t)}\frac{P_0(t)}{P_1(t)}\norm{C\psi^0(t)}^2,&\, 
    \Delta(t) < 0.
  \end{cases}
\end{align}
Inserting the rates~\eqref{eq:TLA_RATEPOS} and \eqref{eq:TLA_RATENEG}  as well
as the analytical solutions of Appendix~\ref{app:TLA} for the probabilities 
$P_0(t)$ and $P_1(t)$ into 
Eq.~\eqref{eq:WTD_ODE}, 
we may solve a formal expression for the WTD. 
The solution depends on the particular path
that one realization might take. For example, 
WTD is different for a jump $\psi^0\to\psi^1$
somewhere in the interval $[T,T+\tau]$  
if the realization has made zero or two transitions before 
time $T$. We illustrate this in Fig.~\ref{fig:TLA},
where we have plotted the decay rate $\Delta(t)$, three sample 
realizations and the WTDs for each realization solved 
from Eq.~\eqref{eq:WTD_ODE} and also from Eq.~\eqref{eq:WTD_EST}.
The initial state is $\psi^0(0)=\ket{0}$ and we use parameter values
$\gamma_0=5\lambda$, 
$\delta=8\lambda$ (see Appendix~\ref{app:TLA}) and a sample size of $10^5$. 
Points of discontinuity in the waiting time distribution in panel 
e) of Fig.~\ref{fig:TLA} correspond
to jumps and since the state of the realization changes, the waiting time 
distribution also changes. We see
that during periods of negative decay rate, the derivative of  
WTD is zero for realizations that are in state 
$\psi^0$, since
the jump rate is zero. 

This system is the simplest one since it has only 
one decay channel and the pure state decomposition of 
Eq.~\eqref{eq:decomp} consists of two states. Jump paths between the different
states in the pure state decomposition show that, 
both the positive and the negative channel act as a one-to-one 
map in the projective Hilbert space of the system.

It is interesting to consider the
WTD for  
a realization, which jumps at some time during 
the first positive decay rate region and then makes a reverse 
jump during the first negative region.
For the first positive region $[t_0,t_1)$ we obtain
\begin{align}
\label{eq:1} 
 F(\tau|\psi^0,t_0)=&\frac{\norm{\tilde\psi^0_{t_0}(0)}^2
    -\norm{\tilde\psi^0_{t_0}(\tau)}^2}{\norm{\tilde\psi^0_{t_0}(0)}^2},
\end{align}
and for the first negative region $[t_1,t_2)$
\begin{align}
\label{eq:2}
  F(\tau|\psi^1,t_1)=&\frac{\norm{\tilde\psi^0_{t_0}(t_1+\tau-t_0)}^2
    -\norm{\tilde\psi^0_{t_0}(t_1-t_0)}^2}{1-\norm{\tilde\psi^0_{t_0}(t_1-t_0)}^2}.
\end{align}
When comparing the WTD of Eq.~\eqref{eq:1} for the positive jumps to
the WTD of Eq.~\eqref{eq:2} for the negative
jumps, we see that they are complementary: in the numerator 
the norm decrease of the state $\tilde{\psi}^0$ in the positive region 
is switched to a norm increase in the negative region
and, the denominator in the negative region is the complement of the denominator 
in the positive region. 

Equations~\eqref{eq:1} and~\eqref{eq:2} provide a simple way of doing a simulation 
for the TLA. For example, during the $k$th negative region we could 
calculate the random 
waiting time using Eq.~\eqref{eq:2} with substitution $t_1\to t_k$, 
for each realization that are in the ground state. During the $k$th 
negative period realizations that are not in the ground state do not 
have a possibility to jump. During the $k$th positive period we would 
use Eq.~\eqref{eq:1} with $t_0\to t_{k-1}$ for jumps away from the state 
$\psi^0$.
  
\subsection{$\Lambda$-system}\label{sec:LAMBDA}
Let us indicate the basis for the Hilbert space of the system  
with $\left\{\ket{0},\ket{1},\ket{2}\right\}$,
where $\ket{1}$ and $\ket{2}$ are the ground states and $\ket{0}$ is  
the common excited state. 
Schematic representation of this system is in Fig.~\ref{fig:systems}b).
Initial state is 
$\psi^0(t_0)=c_0(t_0)\ket{0}+c_1(t_0)\ket{1}+c_2(t_0)\ket{2}$,
which is the only state with non-trivial deterministic evolution (see
Appendix~\ref{app:LAMBDA} for more details). 

Deterministic evolution is generated by 
$\Heff(t)$ (see Eq.~\eqref{eq:HEFF} and Appendix~\ref{app:LAMBDA}). 
The state of the system $\rho(t)$ can be decomposed for all times $t$ 
as $\rho(t)=\sum_{k=0}^2P_k(t)\op{\psi^k}{\psi^k}$.
States $\psi^k\equiv\ket{k}$, with $k=1,2$, are the ground states 
of the system. The probabilities appearing in the decomposition are explicitly calculated in 
the Appendix~\ref{app:LAMBDA}.

Since we have two decay rates, we have four possible combinations
of the decay rate signs. For each pure state of the 
decomposition we only present the decay rate sign combinations which lead
to a non-zero jump rate away from the state under consideration. 
Other sign combinations would
produce zero rate. Jump rate away from the state $\psi^0$ is
\begin{align}\label{eq:LAMBDA_DET}
  \Gamma[\psi^0,t]=&\begin{cases}
    \sum_i\Delta_i(t)\norm{C_i\psi^0(t)}^2,& \, 
    \Delta_1,\,\Delta_2 \geq 0,\\
    \Delta_i(t)\norm{C_i\psi^0(t)}^2,&\, 
    \Delta_i\geq 0 \wedge \Delta_j<0. 
  \end{cases}
\end{align}
Jump rate away from the ground state $\psi^k$ is
\begin{align}\label{eq:LAMBDA_GROUND}
  \Gamma[\psi^k,t]=&\abs{\Delta_k(t)}\frac{P_0(t)}{P_k(t)}
  \norm{C_k\psi^0(t)}^2,
\end{align}
when $\Delta_k(t)<0$. This is the case irrespective of the sign
of the other decay rate.

Both channels, irrespective of the sign of the decay rate, are one-to-one maps.
However, when both channels are positive, $\psi^0$ may be mapped 
to $\psi^1$ or $\psi^2$ when considering the effect of 
both channels.
All rates are proportional to $\norm{C_k\psi^0(t)}^2=\abs{c_0(t)}^2$.

\begin{figure}
  \includegraphics{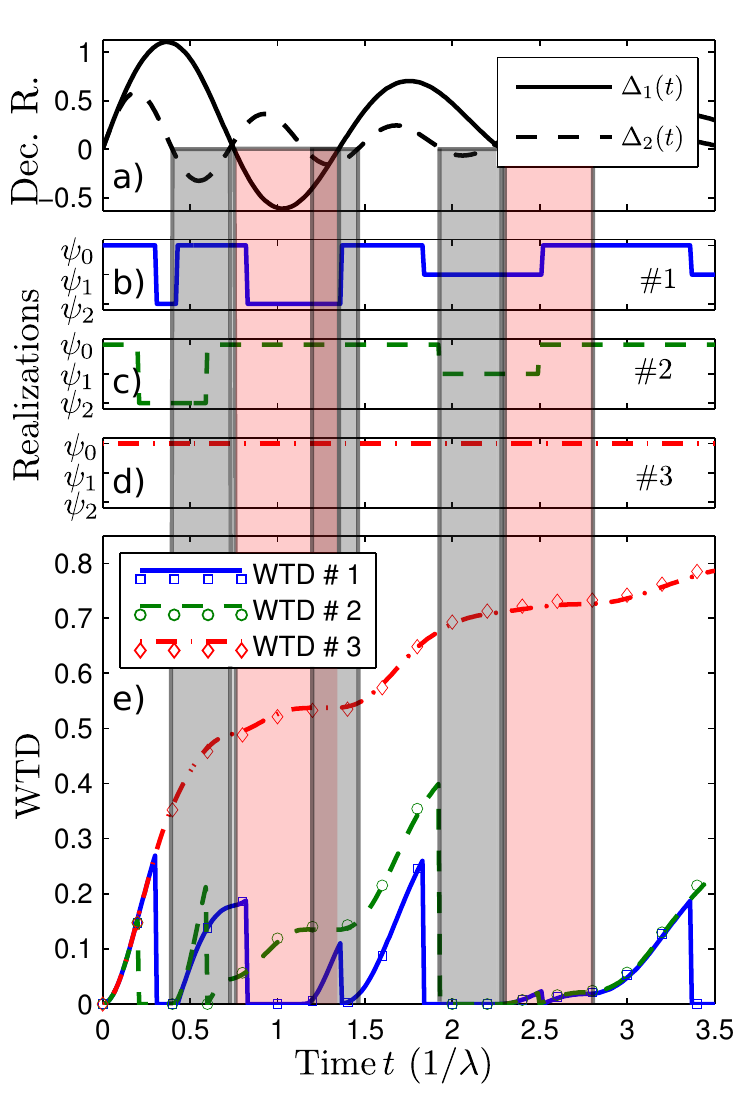}
  \caption{\label{fig:LAMBDA}(Color online)
  Initial state is $\ket{\psi^0(0)}=\ket{0}$ and parameters are 
  $\gamma_0^{(1,2)}=5\lambda$, 
  $\delta^{(1)}=4\lambda$, $\delta^{(2)}=8\lambda$ 
  and sample size was $10^5$. In panel $a)$
  we have the decay rate, in $b)-d)$ we have three 
  different realizations. In panel $e)$ we have 
  WTDs for the realizations. Line style and color coding match with
  the sample realization. Lines are for exact numerical solution and markers
  for sample estimate.}
\end{figure}

Some realizations are plotted together with their  WTD 
in Fig.~\ref{fig:LAMBDA}. The initial state is $\psi^0(0)=\ket{0}$ and
we use parameter values 
$\gamma_0^{(1,2)}=5\lambda$, $\delta^{(1)}=4\lambda$, 
$\delta^{(2)}=8\lambda$ (see Appendix~\ref{app:LAMBDA}) 
and a sample size of $10^5$.
We obtain an interesting expression for the WTD for a reverse jump
$\psi^1\to \psi^0$, if we let
$\Delta_1(t)<0$ during time intervals $[s_1^1,s_2^1]$, $[s_3^1,s_4^1]$, etc. If 
a jump to the state $\psi^1$ occurred at time $T\in[t_0,s_1^1]$ then 
the probability for jump away from $\psi^1$ somewhere in 
the interval $[t_0,T+\tau]$, where $T+\tau\in[s_{2n-1}^1,s_{2n}^1]$, is  
\begin{align}
  F(\tau|\psi^1,T)=1-\frac{P_1(s_2^1)}{P_1(s_1^1)}
  \frac{P_1(s_4^1)}{P_1(s_3^1)}\cdots
  \frac{P_1(T+\tau)}{P_1(s_{2n-1}^1)}.
\end{align}
Since $\Delta_1(t)<0$ when $t\in[s_{2n-1}^1,s_{2n}^1]$, the
probabilities $P_1(s_{2n}^1)<P_1(s_{2n-1}^1)$. Therefore, 
each fraction is smaller than unity  and $F(\tau|\psi^1,T)$ 
is monotonically increasing function.  
\subsection{Ladder system}\label{sec:ladder-system}
We label the orthonormal basis for the Hilbert space of the system
with $\left\{\ket{0},\ket{1},\ket{2}\right\}$,
where $\ket{0}$ is the excited state, $\ket{1}$ is the middle state and
$\ket{2}$ is the ground state. 
Schematic representation of this system is in Fig.~\ref{fig:systems}c).
The initial state is of the form 
$\psi^0(t_0)=c_0(t_0)\ket{0}+c_1(t_0)\ket{1}+c_2(t_0)\ket{2}$.
The deterministic evolution is generated by
$\Heff(t)$ (see Appendix~\ref{app:LADDER} and Eq.~\eqref{eq:HEFF}). 
For all times $t$ the state of the system $\rho(t)$ 
may be decomposed as 
$\rho(t)=\sum_{k=0}^2P_k(t)\op{\psi^k}{\psi^k}$, where
$\psi^k=\ket{k}$, with $k=1,2$, are the middle and the ground states, 
respectively.
Analytical expressions for the probabilities
$P_i(t)$ are in Appendix~\ref{app:LADDER}. For this system, the only state 
invariant in respect to $\Heff$ is $\psi^2$ (see Appendix~\ref{app:LADDER}).

As in Sec.~\ref{sec:LAMBDA} we write down only those combinations 
of the decay rates that give a non-zero jump rate. For the 
initial state 
$\psi^0(t)$ we have
\begin{align}
  \Gamma[\psi^0,t]=&\begin{cases}
    \sum_k\Delta_k(t)\norm{C_k\psi^0(t)}^2,&\, \Delta_1,\Delta_2 \geq 0,\\
    \Delta_i(t)\norm{C_i\psi^0(t)}^2,&\, \Delta_i\geq 0 \wedge \Delta_j<0,
  \end{cases}
\end{align}
and for the middle state $\psi^1(t)$ we have
\begin{align}
  \Gamma[\psi^1,t]=&\begin{cases}
    \Delta_2(t),&\, \Delta_1,\Delta_2 \geq 0,\\
    \Delta_2(t)\\
    \,+\abs{\Delta_1(t)}\frac{P_0(t)}{P_1(t)}\norm{C_1\psi^0(t)}^2,&\,
    \Delta_2\geq 0 \wedge \Delta_1<0,\\
    \abs{\Delta_1(t)}\frac{P_0(t)}{P_1(t)}\norm{C_1\psi^0(t)}^2,&\,
    \Delta_1,\Delta_2 <0,
  \end{cases}
\end{align}
and for the ground state $\psi^2$ we have 
\begin{align}
\label{eq:3}
  \Gamma[\psi^2,t]=&\abs{\Delta_2(t)}\left(
  \frac{P_0(t)}{P_2(t)}\norm{C_2\psi^0(t)}^2+\frac{P_1(t)}{P_2(t)}\right),
\end{align}
when $\Delta_2(t)<0$ irrespective of the sign of $\Delta_1(t)$.

Channel $1$ maps $\psi^0$ to $\psi^1$ and channel $2$
maps $\psi^0$ to  $\psi^2$ and $\psi^1$ to $\psi^2$.
When decay rates are negative, channel $1$ maps 
$\psi^1$ to $\psi^0$. However, channel $2$ maps 
$\psi^2$ to $\psi^1$ or $\psi^0$ when negative. Therefore, when jump to
channel 2 occurs when it is negative, we still have a probability distribution
over the two different target states from which, we have to choose the 
actual target state for the jump.

\begin{figure}
  \includegraphics{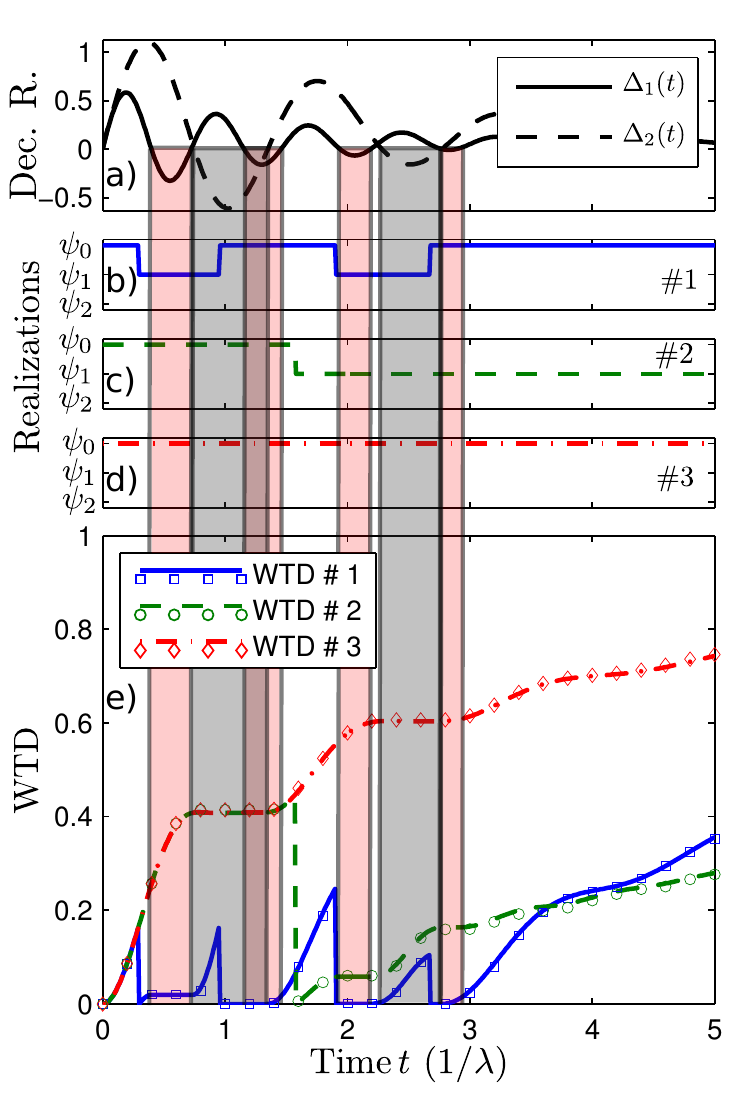}
  \caption{\label{fig:LADDER}(Color online)
  Initial state is  $\ket{\psi^0(0)}=\ket{0}$ and parameters are 
  $\gamma_0^{(1,2)}=5\lambda$, 
  $\delta^{(1)}=8\lambda$, $\delta^{(2)}=4\lambda$ 
  and sample size was $10^6$. In panel $a)$
  we have the decay rate, in $b)-d)$ we have three 
  different realizations. In panel $e)$ we have 
  WTDs for the realizations. Line style and color coding match with
  the sample realizations. Lines are for exact numerical solution and markers
  for sample estimate.}
\end{figure}

In Fig.~\ref{fig:LADDER} we have plotted decay rates and three realizations 
with their respective WTD. There, 
the initial state we use is $\psi^0(0)=\ket{0}$ and the parameters are  
$\gamma_0^{(1,2)}=5\lambda$, $\delta^{(1)}=8\lambda$, 
$\delta^{(2)}=4\lambda$ (see Appendix~\ref{app:LADDER}) and we used a 
sample size of $10^6$.

It has been shown in Ref.~\cite{Piilo2009} that for some parameter values
the approximations made while obtaining the master equation for
this level geometry fail, which is manifested by the breakdown of positivity. 
This is due to the fact that the population of the  
ground state $\psi^2$ is drained completely while the decay rate 
$\Delta_2(t)$ is 
still negative. This causes Eq.~\eqref{eq:3} to diverge. 
Let us assume that $\Delta_2(t)<0$ during intervals the
$[t_1^2,t_2^2]$, $[t_3^2,t_4^2]$, etc. and that at time $T\in [t_0,t_1^2)$
the realization jumps to state $\psi^2$. Assuming that 
$T+\tau\in[t_{2n-1}^2,t_{2n}^2]$ then, the analytical form for
WTD reads
\begin{align}
  \label{eq:4}
  F(\tau|\psi^2,T)=& 1-\frac{P_2(t_2^2)}{P_2(t_1^2)}\frac{P_2(t_4^2)}{P_2(t_3^2)}
  \cdots\frac{P_2(T+\tau)}{P_2(t_{2n-1}^2)}.
\end{align}
From Eq.~\eqref{eq:4} we see that if $\lim_{t'\to T+\tau}P_2(t')=0$,
waiting time distribution reaches unity in finite time but it is 
still well defined. Dynamical consequences of this are that 
a simulation method utilizing full WTD would not break down. Instead, population
of the state $\psi^2$ would go to zero and the total population 
is distributed between the pure states $\psi^0$ and $\psi^1$.

\section{Discussion}\label{sec:DISCUSS}
The positive definite jump probability density of 
Eq.~\eqref{eq:conditional_jump_density} shows that
there is a correlation between the different regions of the projective
Hilbert space. 
Therefore, a general form of the WTD in Eq.~\eqref{eq:WTD_GEN} is complicated since 
it takes the correlation into  account cumulatively. On the other hand, it confirms 
that the realizations of the PDP considered in this paper do not form a trajectory, 
i.e. continuous measurement interpretation can not be necessarily made.  This happens 
because it is not possible to express the WTD for a given realization in terms of that 
particular realization only. This is the argument used already by Gambetta and Wiseman 
in the context of non-Markovian quantum state
diffusion~\cite{PhysRevA.66.012108} but it can be also applied here. For further 
discussion on this highly non-trivial topic, we refer the reader to 
Refs.~\cite{PhysRevLett.100.080401,PhysRevLett.101.140401,PhysRevLett.101.149902}.
        
In the case that there is a state $\psi^k$ in the 
pure state decomposition of $\rho(t)$ 
that acts only as a source state for jumps for 
some period $[T,T+\tau]$, then the WTD is quite simple over this period. 
During this period, the probability of the state $\psi^k$ in 
the pure state decomposition
changes only by jumps away from that state. Hence, we have the 
following identity
$P_k(T+\tau)=P_k(T)-F(\tau|\psi^k,T)P_k(T)$ from which 
we can solve  
\begin{align}\label{eq:WTD_SIMPLE}
  F(\tau|\psi^k,T)=& \frac{P_k(T+\tau)-P_k(T)}{P_k(T)},
\end{align} 
where $P_k(T)\neq 0$ is assumed.
In the examples that we considered in this work this happens 
in the TLA always; in the $\Lambda$-system always for ground states and for 
the state $\psi^0$, when the decay rates have the same signs; and in 
the Ladder system for the ground state always, for the middle state 
when the decay rates have the opposite signs and for the state $\psi^0$
when the decay rates rates have equal signs.

For a short time interval 
$\delta t$ we can approximate the full WTD as
\begin{align}
  F(\delta t|\psi,T)\approx\Gamma[\psi(T),T]\delta t=
  \int\text{d}\phi\,\sum_kp_k[\phi|\psi].
\end{align} 
Thus, for a short time interval  
the total jump rate is resolvable in (channel, target state) pairs: each
channel maps a source state to a target state (one-to-one relation 
for the Markovian jumps
and one-to-many for the non-Markovian jumps).
During this short interval,
the occurrence of a jump excludes the possibility of another jump at the 
same interval to another channel. 
In WTD-based methods 
these individual contributions are cumulatively gathered together. The process 
may be reset after any time interval $\Delta t$, after which a new 
random number must be drawn. 
In the limit $\Delta t\to \delta t$, stepwise method 
emerges.  

\section{Conclusions}
\label{sec:CONCLUSIONS}
We have derived a general waiting time distribution of quantum jumps for open 
quantum systems following non-Markovian dynamics. In this sense, our results 
generalize the QT methods into the non-Markovian regime. The distribution is 
a well defined conditional probability distribution function which takes 
into account in a proper manner the
bidirectional probability flow between different regions of the
projective Hilbert space of the system.
The WTD includes probabilities which are
present in the pure state decomposition of the reduced system state, i.e., 
the realizations of the process depend on each other -- a feature stemming 
from the memory effects and present already in the NMQJ method.
Our results seem to confirm the view that the realizations of the PDP, that 
the WTD govern, 
do not form a trajectory, therefore the PDP can not be interpreted 
in terms of a continuous measurement of the environment. 
We have constructed the WTD explicitly for some quantum optical systems and 
also discussed the cases when the calculation of the WTD can be simplified. 

Our work complements the theory of Monte Carlo methods for non-Markovian 
systems and the WTD concept familiar from Markovian regime is now 
also well-defined for non-Markovian systems. We hope that this work stimulates 
further research for 
non-Markovian dynamics and especially inspires 
new directions in the development of simulation tools for open quantum systems.

\begin{acknowledgments}
The authors would like to thank Academy of Finland (projects 133682 and 
259827), COST Action MP1006, Jenny and Antti Wihuri Foundation, Magnus 
Ehrnrooth Foundation, and Vilho, Yrj\"o and Kalle V\"ais\"al\"a
Foundation for financial support and Sascha Wallentowitz for stimulating 
discussions.
\end{acknowledgments}
\appendix

\section{Markovian limit}\label{sec:markovian-limit}
In the Markovian limit, decay rates $\Delta_i(t)\geq 0$, for all times $t$.
Then, the jump rate away from state $\psi$ at time $T+s$ is
\begin{align}\label{eq:7}
  \Gamma[\psi,T+s]=&\sum_i\Delta_i(T+s)\norm{C_i\psi(T+s)}^2\notag\\
  =&
  \sum_i\Delta_i(T+s)\frac{\norm{C_i\tilde{\psi}_T(s)}^2}{\norm{\tilde{\psi}_T(s)}^2},
\end{align}
where we have used notation of Eq.~\eqref{eq:8}. On the other hand,
\begin{align}\label{eq:9}
 \frac{\text{d}}{\text{d}s}\norm{\tilde{\psi}_T(s)}^2=&-\sum_i\Delta_i(T+s)\norm{C_i\tilde{\psi}_T(s)}^2,
\end{align}
where we have used Schr\"odinger equation and Eq.~\eqref{eq:HEFF}. Therefore, the total
jump rate is
\begin{align}\label{eq:10}
  \Gamma[\psi,T+s]=&-\frac{\frac{\text{d}}{\text{d}s}\norm{\tilde{\psi}_T(s)}^2}{\norm{\tilde{\psi}_T(s)}^2}=
  -\frac{\text{d}}{\text{d}s}\ln{\norm{\tilde{\psi}_T(s)}^2}.
\end{align}
Now, using Eq.~\eqref{eq:WTD_IMPLICIT} we obtain 
\begin{align}\label{eq:11}
  F(\tau|\psi,T)=&1-\exp\left\{\int_0^\tau\text{d}s\,\ln{\norm{\tilde{\psi}_T(s)}}\right\}\notag\\
  =&1-\exp\left\{\ln\left\{\frac{\norm{\tilde{\psi}_T(\tau)}^2}{\norm{\tilde{\psi}_T(0)}^2}\right\}\right\}\notag\\
  =&\frac{\norm{\tilde{\psi}_T(0)}^2-\norm{\tilde{\psi}_T(\tau)}^2}{\norm{\tilde{\psi}_T(0)}}.
\end{align}
\section{System definitions}
We are considering two- and three level atoms interacting with a leaky
cavity mode. The spectral density of the cavity is
\begin{align}\label{eq:6}
  J(\omega)=\frac{1}{2\pi}\frac{\gamma_0\lambda^2}
{(\omega-\omega_c)^2+\lambda^2},
\end{align}
where $\gamma_0$ is the coupling constant, $\lambda$ is the width of the 
Lorentzian and $\omega_c$ is the cavity resonance frequency. Another 
important parameter is the detuning of the atom from the
cavity resonance: $\delta=\omega_a-\omega_c$, where $\omega_a$ is one 
of the transition frequencies of the atom. 

The time convolutionless master equations for the example systems in 
Sec.~\ref{sec:EXAMPLES}, are all special cases from the following 
general form
\begin{align}\label{eq:Master2}
\dot{\rho}(t)&=-i\left[\sum_ks_k(t)C_k^\dagger C_k,\rho(t)\right]\notag\\
&+
\sum_k\Delta_{k}(t)\left(C_k\rho(t)C_k^\dagger-
\frac{1}{2}\left\{\rho(t),C_k^\dagger C_k\right\}\right),
\end{align}
where $s_k(t)$ is the time dependent Lamb shift, $\Delta_k(t)$ are 
the time dependent decay rates and $C_k$ are the time independent
jump operators. For simplicity, we have assumed in the actual calculations that
$s_k(t)=0$. 
We use TCL4 
approximated analytical form 
for the decay rate \cite{Breuer2007} corresponding to a
spectral density of Eq.~\eqref{eq:6}
\begin{align}
  \Delta(t)=&
  \frac{\gamma_0\lambda^2}{\lambda^2+\delta^2}\left[1-e^{-\lambda t}
    \left(\cos(\delta t)-\frac{\delta}{\lambda}\sin(\delta t)\right)\right]
  \notag\\
  &+\frac{\gamma_0^2\lambda^5e^{-\lambda t}}{2\left(\lambda^2+\delta^2\right)^3}
  \Bigg\{\left[1-3\left(\frac{\delta}{\lambda}\right)^2\right]\left(
    e^{\lambda t}-e^{-\lambda t}\cos(2\delta t)\right)\notag\\
    &-2\left[1-\left(\frac{\delta}{\lambda}\right)^4\right]\lambda t 
    \cos(\delta t)+4\left[1+\left(\frac{\delta}{\lambda}\right)^2\right]
    \delta t\sin(\delta t)\notag\\
    &+\frac{\delta}{\lambda}\left[3-\left(\frac{\delta}{\lambda}\right)^2\right]
    e^{-\lambda t}\sin(2\delta t)\Bigg\}.
\end{align}

\subsection{Two level system}\label{app:TLA}
Master equation:
\begin{align}
  \dot{\rho}(t)=&-is(t)[\op{0}{0},\rho(t)]+\Delta(t)\op{1}{0}\rho(t)\op{0}{1}
  \notag\\
  &-\frac{1}{2}\Delta(t)\left\{\rho(t),\op{0}{0}\right\}.
\end{align}
Jump operator:
\begin{align}
  C=\op{1}{0}.
\end{align}
Solution for the probabilities in the pure state decomposition of
Sec.~\ref{sec:TLA}:
\begin{align}
  P_0(t)=&\norm{\tilde\psi_{t_0}^0(t-t_0)}^2,\\
  P_1(t)=&1-\norm{\tilde\psi_{t_0}^0(t-t_0)}^2.
\end{align}
\subsection{$\Lambda$-system}\label{app:LAMBDA}
Master equation:
\begin{align}
  \dot{\rho}(t)=&-is_1(t)[\op{0}{0},\rho(t)]-is_2(t)[\op{0}{0},\rho(t)]\notag\\
  &+\Delta_1(t)\left[\op{1}{0}\rho(t)\op{0}{1}
    -\frac{1}{2}\{\rho(t),\op{0}{0}\}\right]\notag\\
  &+\Delta_2(t)\left[\op{2}{0}\rho(t)\op{0}{2}
    -\frac{1}{2}\{\rho(t),\op{0}{0}\}\right].
\end{align}
Jump operators:
\begin{align}
  C_1=&\op{1}{0},\\
  C_2=&\op{2}{0}.
\end{align}
Solution for the probabilities in the pure state decomposition of 
Sec.~\ref{sec:LAMBDA}:
\begin{align}
  P_0(t)=&\norm{\tilde\psi^0_{t_0}(t-t_0)}^2,\notag\\  
  P_j(t)=&\int_{t_0}^t\text{d}s\,\Delta_j(s)\abs{\tilde{c}_0(s)}^2, 
\end{align}
for $j\in\{1,2\}$.
\subsection{Ladder system}\label{app:LADDER}
Master equation:
\begin{align}
  \dot{\rho}(t)=&-is_1(t)[\op{0}{0},\rho(t)]-is_2(t)[\op{1}{1},\rho(t)]\notag\\
  &+\Delta_1(t)\left[\op{1}{0}\rho(t)\op{0}{1}
    -\frac{1}{2}\{\rho(t),\op{0}{0}\}\right]\notag\\
  &+\Delta_2(t)\left[\op{2}{1}\rho(t)\op{1}{2}
    -\frac{1}{2}\{\rho(t),\op{1}{1}\}\right].
\end{align}
Jump operators:
\begin{align}
  C_1=\op{1}{0},\\
  C_2=\op{2}{1}.
\end{align}
Solution for the probabilities in pure state decomposition of 
Sec.~\ref{sec:ladder-system}:
\begin{align}
P_0(t)=&\norm{\tilde\psi^0_{t_0}(t-t_0)}^2,\\ 
P_1(t)=&e^{-D_2(t)}\int_{t_0}^t\text{d}s\,\Delta_1(s)e^{-D_1(s)+D_2(s)}
\norm{C_1\psi^0(t_0)}^2,\\ 
P_2(t)=&\left(1-e^{-D_2(t)}\right)\norm{C_2\psi^0(t_0)}^2
+\int_{t_0}^t\text{d}s\,\Delta_2(s)P_2(s),
\end{align}
where $D_i(t)=\int_{t_0}^t\text{d}s\,\Delta_i(s)$.
\bibliography{manu}
\end{document}